\documentclass[12pt]{article} %latex
\usepackage{cite} %Latex
\usepackage{latexsym}
\usepackage{amsmath}
\usepackage{amssymb}
\usepackage{pstricks}
\usepackage{indentfirst}
\newcommand{\becs}{\begin{cases}}
\newcommand{\bem}{\begin{matrix}}

\newcommand{\dya}[1]{|#1\rgl\lgl#1|}

\newcommand{\encs}{\end{cases}}
\newcommand{\enm}{\end{matrix}}

\newcommand{\hf}{{\textstyle\frac{1}{2} }}

\newcommand{\inpd}[2]{\lgl#1|#2\rgl }
\newcommand{\ket}[1]{|#1\rgl }

\newcommand{\lgl}{\langle } 
\newcommand{\mte}[2]{\lgl#1|#2|#1\rgl }
\newcommand{\mted}[3]{\lgl#1|#2|#3\rgl }
\newcommand{\od}{\odot }
\newcommand{\ot}{\otimes }
\newcommand{\pr}{\partial }
\newcommand{\rgl}{\rangle }

\newcommand{\vb}{\,|\,}

\newcommand{\FS}{{\cal F}}

\newcommand{\bld}[1]{\boldsymbol{#1}}

\newcommand{\al}{\alpha }
\newcommand{\bt}{\beta }

\newcommand{\dl}{\delta }

\newcommand{\om}{\omega }

\newcommand{\onlinecite}[1]{\citen{#1}} %Provides Revtex equivalent
\renewcommand{\cite}[1]{$^{\citen{#1}}$} %Puts citation in superscript

 \newcommand{\lbrc}{\left\{\rule{0cm}{2.5ex}\right.}
 \newcommand{\csp}{@{\hspace{0.2em}}} 
 \newcommand{\dsp}{@{\hspace{0.05em}}}
 \newcommand{\rbx}[1]{\raisebox{1.5 ex}[0cm][0cm]{$#1$}}

\begin{document}

\title{Quantum Mechanics Without Measurements}

%latex version:
\author{Robert B. Griffiths\\
Physics Department,
Carnegie-Mellon University,\\
Pittsburgh, PA 15213, USA}

%revtex version:
%\author{Robert B. Griffiths}
%\email{rgrif@cmu.edu}
%\affiliation{%
%Department of Physics,
%Carnegie-Mellon University,\\
%Pittsburgh, PA 15213, USA}

\date{Version of 7 December 2006}

\maketitle  % latex

\begin{abstract}

Many of the conceptual problems students have in understanding quantum
mechanics arise from the way probabilities are introduced in standard
(textbook) quantum theory through the use of measurements.  Introducing
consistent microscopic probabilities in quantum theory requires setting up
appropriate sample spaces taking proper account of quantum
incompatibility. When this is done the Schr\"odinger equation can be used to
calculate probabilities independent of whether a system is or is not being
measured, and the results usually ascribed to wave function collapse are
obtained in a less misleading way through conditional probabilities. Toy models
that include measurement apparatus as part of the total quantum system make
this approach accessible to students.  Some comments are made about teaching
this material.

\end{abstract}

%\maketitle  %revtex

	\section{Introduction}
\label{sct1}

Quantum mechanics is a difficult subject to teach, and there has been a
significant effort to find out what problems students have in understanding it,
and how to overcome them.\cite{ctk01,br02,sbc06}.  In part the difficulties
arise from unfamiliar mathematics: partial differential equations, complex
linear algebra (or functional analysis), and probability theory.  However, the
greatest difficulty is surely the one encapsulated in Feynman's well-known
assertion that ``Nobody understands quantum mechanics.''\cite{fn64} How are
students to learn a subject that their teachers do not understand?

Feynman's own masterful exposition of the subject\cite{fn63} is proof that
physicists can, indeed, teach what we do not understand, or do not understand
as well as we would like to.  At the same time, what he said needs to be taken
seriously; Feynman was not joking.  The problems he had understanding the
subject are also severe barriers to less brilliant minds, and the basic thesis
of this article is that helping students overcome them, rather than sweeping
them under the carpet, is well worth the effort.  That is most obvious in the
case of future professional physicists or electrical engineers who will need to
deal with entanglement, quantum information, transport in nanocircuits, and
similar subjects for which the approach found in current textbooks does not
provide a helpful physical intuition.  But for the sake of other students as
well, we need to try and counter the quasi-magical view of the quantum world
that results from trying to make sense of what one finds in current textbooks,
not to mention popular expositions written by authors who understand quantum
mechanics even less that Feynman did, but are less forthright in confessing
their ignorance.

At the heart of the conceptual difficulties of quantum theory is the failure of
the current textbook version of the subject---often called ``Copenhagen'' or
``standard'' quantum me\-cha\-nics---to introduce probabilities in quantum
theory in a consistent and meaningful way.  Instead, probabilities are
introduced on the basis of measurements, an approach which conveniently gets
around various difficulties, but leaves students with a confused idea of what
quantum mechanics is all about, and the impression that understanding the
subject is impossible.  Instead they get the feeling that one should, to use
Mermin's memorable phrase,\cite{mr04} ``shut up and calculate.''  That
measurements provide an unsatisfactory approach to quantum interpretation has
been known for a long time in the quantum foundations
community,\cite{wz83,ctk20} where the ``measurement problem'' is widely
considered both an embarrassment\cite{bl90} and an intractable
difficulty\cite{mt98}.  More about this in Sec.~\ref{sct2}.

The consistent or decoherent histories, hereafter abbreviated to ``histories,''
approach to quantum theory\cite{ctk10,ctk11,ctk12,gr02} allows one to
introduce probabilities in a physically meaningful and mathematically
consistent way without reference to measurements.  Doing so requires that one
confront head on the central conceptual difficulty of quantum mechanics:
quantum incompatibility.  This is discussed in Sec.~\ref{sct3} in terms of a
spin-half particle. No interpretation of quantum mechanics can be considered
satisfactory if it cannot make both mathematical (the easy part) and physical
(the hard part) sense of this, the simplest of quantum systems.

Probabilistic or stochastic time development in quantum mechanics requires the
notion of a quantum history, a concept which in itself is not particularly
difficult, Sec.~\ref{sct4}.  Assigning probabilities without using measurements
can then be done using the Born rule (not hard) and its extensions
(more subtle) applied to a closed or isolated quantum system, i.e., one to
which Schr\"odinger's equation applies.  What is going on in a real measurement
process using quantum mechanical apparatus (no other kind is currently
available) can then be understood by applying the fundamental probabilistic
laws of quantum mechanics to the measured system and apparatus, regarded as
a single quantum system.   The discussion in Sec.~\ref{sct4} attempts to
communicate the essential ideas while omitting the technical machinery that is
available elsewhere.\cite{ntk10}

In addition to the rules, students need simple examples which illustrate in
physical terms what the formalism is all about.  Section~\ref{sct5} is a brief
introduction to what I call ``toy models,'' with application to a decaying
nucleus and the subsequent detection of an alpha particle.  This shows how
quantum theory can be applied in principle to analyze real measurements without
treating ``measurement'' as an axiom, and without using wave function collapse.
The topic of measurements continues in Sec.~\ref{sct6}, where it is explained
how and why one can interpret a Stern-Gerlach measurement as revealing a value
of spin angular momentum \emph{before} the particle was measured, and why the
usual textbook approach, though not wrong, is seriously misleading.

All well and good, but can one teach this new understanding to students who are
not as bright as Feynman?  In Sec.~\ref{sct7} I discuss my own experience,
along with a few practical problems involved in introducing the new approach
into the curriculum.  The more difficult question of whether doing so is
worthwhile is taken up in the concluding Sec.~\ref{sct8}.

\section{What Is Wrong With Measurements}
\label{sct2}

The basic difficulty Feynman and everyone else has had understanding quantum
theory comes from the need to introduce probabilities into the theory in a
consistent way.  It is well known that Einstein was opposed on aesthetic and
philosophical grounds to a theory that was random at the fundamental level: God
does not throw dice. However, his search for a deterministic quantum mechanics
ended in failure, and at present the prospects of finding such a theory do not
look hopeful.  Consider the decay of radioactive nuclei, such as carbon 14.  So
far as we know at present, this is a purely random process: there is nothing
inside a particular carbon 14 nucleus which determines whether it will decay 10
minutes from now, or in 10 years or in 10,000 years.  No experiment has been
able to separate any species of nucleus of this sort into a batch that will
decay quickly and one that will take longer.\cite{ntk01} The simplest
explanation is that there is nothing in the nucleus before it decays, no
``hidden variable,'' that determines when it will decay.  Attempts to introduce
hidden variables into quantum mechanics lead both to a more complicated theory,
and as shown by Bell,\cite{bl64} to mysterious long-range influences for which
there is not the slightest experimental evidence.\cite{ntk02} Thus it seems
that most contemporary physicists have abandoned Einstein's hope for a
deterministic theory and accept the need to understand quantum mechanics as
something intrinsically probabilistic or stochastic, as first proposed by
Born\cite{br26} in 1926, shortly after Schr\"odinger published his famous
(time-dependent) equation.

But how to introduce probabilities into quantum theory?  The textbook approach
employs \emph{measurements}, and if the textbook has been carefully written
these probabilities refer to \emph{measurement outcomes}, traditionally called
``pointer positions'' in the quantum foundations literature, and \emph{not} to
the microscopic events the apparatus was designed to measure.  There is a very
good reason for making this distinction. The naive assumption that every
conceivable measurement outcome corresponds to a microscopic property leads to
many paradoxes.\cite{rd87}.  By not talking about what is really being measured
and confining the discussion to the macroscopic world, where classical physics
applies to a good approximation, the paradoxes are avoided, and one has a
consistent way of handling experimental results stored in macroscopic form in
photographs or on magnetic disks.  This ``black box'' approach, in which
quantum wave functions and density operators are nothing but mathematical tools
summarizing macroscopic preparation procedures, and used to
calculate probabilities for the outcomes of macroscopic measurements, has much
to recommend it in terms of overall consistency.\cite{dm02}

The trouble with the black box approach is that it provides no physical
intuition about what is going on at the atomic level. Hence the physicist who
wants to understand what the world is all about is no more likely to heed
warnings against opening the box than are the children in one of Grimm's fairy
tales.  While his chances of not getting into trouble are somewhat better than
theirs, he still faces a significant probability of being eaten by the
alligators inhabiting the vast swamp of inconsistent ideas and paradoxes lying
inside the box, or, to change the metaphor, just beneath the surface of
measurement-based interpretations of quantum mechanics.  We know they are there
from many decades of research in quantum foundations.  If teachers and
textbooks cannot bring themselves to be as frank as Feynman, they should at
least consider posting warning signs!  But it would be even better to get rid
of the alligators by draining the swamp: by introducing probabilities for
microscopic events in a fully consistent way, accompanied by an appropriate
physical interpretation.\cite{ntk18}

Yet more confusion is created by treatises that interpret ``measurement'' to
mean a projective measurement of the sort introduced by von Neumann\cite{vn32},
in which a measurement is supposed to ``collapse'' or ``reduce'' the wave
function of the measured system into that eigenstate of the measured
(microscopic) physical variable that corresponds to the apparatus pointer
position.  Most real measurements on microscopic systems are not of this sort.
Far more common are situations in which the measured system is destroyed in the
process of measurement (e.g., a photon is absorbed), or its properties
seriously altered, and the experimentalist interprets the outcome in terms of
properties the measured system had \emph{before} the measurement took place;
e.g., the energy of an alpha particle before it entered and stopped inside a
detector.  Textbook quantum theory thus fails to provide the tools needed to
interpret real experiments in quantum mechanical terms.  In addition, wave
function collapse is a concept which itself gives rise to needless conceptual
headaches. It is not actually needed in quantum theory, since its
real function is that of a tool for calculating conditional probabilities, and
this can be done just as well by other methods which are conceptually clearer
and less likely to mislead; see the example in Sec.~\ref{sct5}.

Rather than treating probabilities as peculiar things somehow associated with
measurements, it is much better to consider them part of the fundamental laws
of nature which apply to \emph{all} quantum phenomena, including measurements
as particular cases.  Why suppose that radioactive nuclei only decay when they
are being measured?  In practice, physicists do not assume that.  Instead, we
calculate decay rates of carbon 14 without asking whether the nuclei are being
measured, or the decay rates of uranium 235 at an epoch when there were no
human beings around to do the measurements, or aluminum 26 in outer space,
where the need to introduce measurements seems even more ludicrous.
Probabilities can, indeed, be introduced as fundamental laws. But before
explaining how to do it, we need to address a central conceptual issue in
quantum theory, which when left unattended leads to all sorts of problems.

\section{Quantum Incompatibility}
\label{sct3}

\begin{figure}[h]
$$
\begin{pspicture}(-6,-2.6)(6,2.3) % (lower left)(upper right)
\newpsobject{showgrid}{psgrid}{subgriddiv=1,griddots=10,gridlabels=6pt}
		%PARAMETERS
\def\lwd{0.035} %Linewidth default
\def\lwb{0.07}  %Broad line
\def\rdot{0.1} \def\rodot{0.2} %Radius of closed, open dot
		%DEFAULTS
%\Huge
\psset{%unit=1.0cm, %1.0 cm is default
labelsep=2.0,
arrowsize=0.150 1,linewidth=\lwd}
		%OBJECTS
	% Dots
\def\dot{\pscircle*(0,0){\rdot}} %Closed dot
\def\odot{\pscircle[fillcolor=white,fillstyle=solid](0,0){\rodot}} %Open dot
	% Cross hatching.  \crhatch{(x1,y1)(x2,y2)... (xn,yn)}
	% NOTE. Set param vals inside [] as desired. 
\def\crhatch#1{\pspolygon[linestyle=none,fillstyle=hlines,hatchangle=30,%
hatchwidth=.02,hatchsep=0.15]#1}
		%DIAGRAM
	\def\classph{
\psellipse[linewidth=\lwb,fillstyle=hlines,hatchangle=30,%
hatchwidth=.02,hatchsep=0.15](0,0)(1.5,1.0)
\psframe[fillcolor=white,fillstyle=solid,linestyle=none](-2,0)(2,1.5)
\psellipse[linewidth=\lwb](0,0)(1.5,1.0)
\psline{->}(-2.5,0)(2.5,0)
\psline{->}(0,-2)(0,2)
\rput[l](2.6,0){$x$}
\rput[b](0,2.1){$p$}
\rput[B](0,-2.5){(a)}
	}
	\def\hilbt{
\psline(-2.5,0)(2.5,0)\psline(0,-2)(0,2)
\psline[linewidth=\lwb](-2.5,-1.25)(2.5,1.25)
\psline[linewidth=\lwb](-1,2)(1,-2)
\psline[linewidth=\lwb](-1.2,-2)(1.2,2)
\rput(1.8,0.9){\dot}\rput(-0.75,1.5){\dot}\rput(0.9,1.5){\dot}
\rput[t](1.8,0.7){$\ket{\psi}$}
\rput[r](-0.9,1.5){$\ket{\tilde\psi}$}
\rput[r](0.7,1.5){$\ket{\chi}$}
\rput[B](0,-2.5){(b)}
	}
\rput(-3.5,0){\classph}
\rput(3.5,0){\hilbt}
%\showgrid %Comment this line to remove grid
\end{pspicture}
$$
\caption{(a) Classical phase space for harmonic oscillator. (b) Two-dimensional
  quantum Hilbert space}
\label{fgr1}
\end{figure}
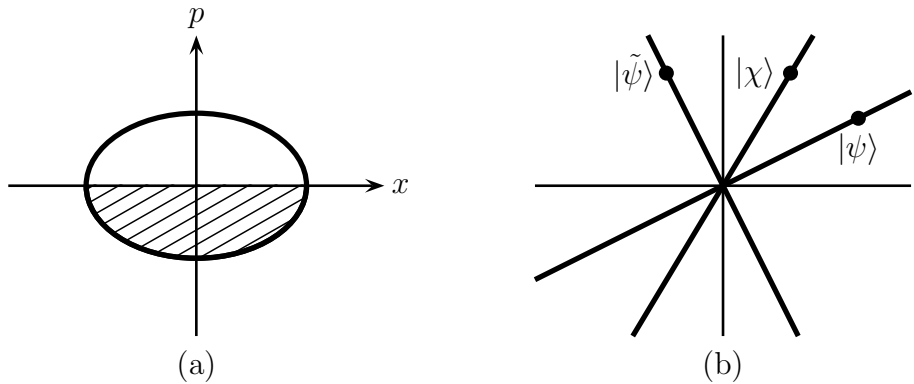

In classical statistical mechanics probabilities are assigned to regions in the
classical phase space.  For a quantum system the analog is the quantum Hilbert
space. However, there is an important differences between the two which needs
to be taken into account in a consistent theory of quantum probabilities.  This
is illustrated in Fig.~\ref{fgr1}, where (a) shows the phase space for a
one-dimensional harmonic oscillator, and (b)---in schematic form, for we have
replaced a complex space with a real space---the two-dimensional Hilbert space
for a spin-half particle.

A \emph{physical property} of a classical particle corresponds to a region in
the phase space where this property is true. For example, the region inside the
ellipse in Fig.~\ref{fgr1}(a) corresponds to the property that the total energy
$E$ is less than $E_0$, while the lower half plane represents the property that
the momentum $p$ is negative. Classical properties combined with
the logical connective AND correspond to the intersection of the corresponding
regions, as in a Venn diagram: $E<E_0$ AND $p<0$ is represented by the shaded
region inside the ellipse. In some cases combining two properties in this way
yields a property which is always false, e.g., $E<E_0$ AND $E> 2E_0$
corresponds to the empty set.  However, it is still meaningful, and the
negation of a false property is a true property.  Negation corresponds
to the set-theoretic complement; thus NOT $E<E_0$ is the same as $E\geq E_0$,
the region outside the ellipse in the figure.

Following von Neumann\cite{ntk05}, we represent a quantum property by a
\emph{ray} or one-dimensional subspace of the Hilbert space, i.e., the
collection of all kets of the form $\{c\ket{\psi}\}$ where $\ket{\psi}$ is
fixed and $c$ is any complex number.  Examples are shown in Fig.~\ref{fgr1}(b).
More generally, a quantum property corresponds to a subspace\cite{ntk06} of the
Hilbert space; e.g., think of a two-dimensional plane passing through the
origin of a three-dimensional space.  The negation of a quantum
property---again we follow von Neumann---is not the set-theoretic complement of
this subspace, but instead its \emph{orthogonal complement}, the subspace of
kets that are orthogonal to (have zero inner product with) all kets in the
original subspace.  Thus in Fig.~\ref{fgr1}(b) the negation of the property
corresponding to $\ket{\psi}$ is represented by the ray $\ket{\tilde\psi}$
perpendicular to the $\ket{\psi}$ ray.  So far as I know, all physicists accept
von Neumann's definition of negation, which makes good physical sense.  For
example, in the case of a quantum harmonic oscillator, $E<E_0$ is naturally
associated with the subspace spanned by linear combinations of energy
eigenstates $\{\ket{n}\}$ which have $(n+\hf)\hbar\om<E_0$, and its negation to
the (infinite-dimensional) subspace spanned by those with $(n+\hf)\hbar\om\ge
E_0$.  In the case of a spin-half particle the negation of $S_z=+1/2$ (in units
of $\hbar$) is the ray corresponding to $S_z=-1/2$.  Since in ordinary logic
either a statement or its negation is true, we conclude that in the case of a
spin-half particle either $S_z=+1/2$ or $S_z=-1/2$, in agreement with the
experimental result of Stern and Gerlach\cite{gs22} when one interprets their
experiment as they themselves did (and which, as we shall see in
Sec.~\ref{sct6}, is fully justified by modern quantum mechanics), as indicating
the property that the particle (in their case a silver atom) had \emph{before}
the measurement took place.

Note, however, a striking contrast between classical and quantum properties,
Fig.~\ref{fgr1}(a) and (b). In the classical case a property and its negation
fill up the entire phase space, whereas for the quantum case the rays
corresponding to $\ket{\psi}$ and its negation $\ket{\tilde\psi}$ do not even
begin to fill up the Hilbert space.  There are plenty of other rays, such as
the one associated with $\ket{\chi}$, a ket which is neither a multiple of
$\ket{\psi}$ nor orthogonal to it.  What can we say about \emph{them}?  In some
sense this is \emph{the} central conceptual difficulty of quantum mechanics,
and no physical interpretation that fails to deal with it---at least no scheme
based on the quantum Hilbert space along with von Neumann's notion of
negation---can hope to succeed.  Von Neumann himself was quite aware of the
problem, and he and Birkhoff\cite{bvn36} in a paper that is at least as
important for the field of quantum foundations as the better known one by
Einstein, Podolsky and Rosen\cite{epr35}, proposed a solution requiring a
radical modification of propositional logic.  Alas, we physicists have not been
able to make much use of it for understanding quantum mechanics. Perhaps we
just are not bright enough, and someday robots will use it to make sense of the
quantum world.  But in the meantime we can make considerable progress using
something much less radical.

The histories approach handles this difficulty through the concept of
\emph{quantum incompatibility}, as per the following illustration.  Whereas
both ``$S_x=+1/2$'' and ``$S_z=-1/2$'' are meaningful statements about a
spin-half particle at a particular instant of time, the logical combination
``$S_x=+1/2$ AND $S_z=-1/2$'' is \emph{meaningless} in the precise sense that
Hilbert-space quantum mechanics can assign it no meaning.  All the rays in the
two-dimensional Hilbert space---at this point we need to think of the complex
analog of Fig.~\ref{fgr1}(b) (points on the Bloch sphere for the reader
familiar with that concept)---already have a physical interpretation, namely
that the component of spin angular momentum in a particular direction in space,
call it $w$, is $+1/2$, and there is none left over which could plausibly
represent ``$S_x=+1/2$ AND $S_z=-1/2$.''  Could this be a statement which is
always false, like the classical $E<E_0$ AND $E> 2E_0$ considered earlier?  The
trouble is that the negation of a meaningful statement which is always false is
one that is always true, such as $E\ge E_0$ OR $E\le 2E_0$ for a classical
oscillator. However, the negation of ``$S_x=+1/2$ AND $S_z=-1/2$,'' which is
``$S_x=-1/2$ OR $S_z=+1/2$,'' does not look like a good candidate for a
statement that is always true, and in fact pursuing this route quickly leads to
contradictory results if one employs the rules of standard logic.\cite{ntk11}
On the other hand, the negation of a meaningless statement is
equally meaningless, so there is no problem as long as we agree that joining
``$S_x=-1/2$'' and ``$S_z=+1/2$'' with OR is no more sensible than joining them
with AND.

Compatibility and incompatibility for larger quantum systems are most
conveniently discussed by considering the projectors (orthogonal projection
operators) onto the subspaces corresponding to the different properties.  If
$P$ and $Q$ are projectors representing two subspaces, or two properties
denoted by the same letters, they are compatible if and only if $PQ=QP$, in
which case $PQ$ is itself a projector onto the subspace corresponding to ``$P$
AND $Q$.'' Otherwise, they are incompatible.  In textbooks the term
``incompatible'' is employed in a similar way, but with reference to
observables (physical variables represented by self-adjoint operators), and one
is told that they cannot be simultaneously measured.  Making reference to
properties (projectors or subspaces) is both technically and conceptually
simpler than referring to observables.  When they are incompatible they indeed
cannot be simultaneously measured, because what is meaningless cannot be
measured.

Since in the classical world everything commutes, there is no exact analog of
quantum incompatibility to be found in our everyday experience.  However, the
following analogies may help tease out some of what it does and does not mean.
A photographer taking pictures of Mount Shasta can do so from a variety of
different directions or perspectives: north, south, east, etc.  The perspective
is chosen by the photographer and has no effect on the reality represented by
the mountain.  The chosen perspective makes it possible to answer certain
questions but not others on the basis of the resulting photograph: a view from
the south will not indicate what is happening on the northern slopes.  
Next, replace the photographer with a classical physicist who has designed an
apparatus to measure the $w$ component of angular momentum of a golf ball by an
apparatus consisting of a cage initially at rest and pivoted on low friction
bearings which allow it to rotate around an axis in the $w$ direction. If it
can be arranged that the moving golf ball flies into and is trapped in the
center of the cage, the final rate of rotation of the cage can be converted
into a value for the angular momentum of the golf ball just before it entered
the cage, i.e., just before the measurement was made.  The choice of
orientation $w$ is made by the physicist, and this choice has no effect upon
the properties of the golf ball prior to the measurement, though it does
determine what he can say about those properties after the measurement is over.
Finally, replace the classical physicist with a quantum physicist who measures
$S_w$ for a spin-half particle using a Stern-Gerlach apparatus with the field
gradient in the direction $w$.  The choice of $w$ is made by the physicist and
has no effect upon the properties of the spin-half particle before it is
measured, a point to which we will return in Sec.~\ref{sct6}.  It does,
however, determine what can be said about the earlier state of the particle
on the basis of the measurement outcome.

How does the last situation differ from the first two? A photographer could
arrange to have a colleague take a picture of Mount Shasta as viewed from the
north at the same time as he takes one from the south, and together the
photographs would provide more information than either one by itself, since the
two perspectives are compatible with each other.  The classical physicist could
in principle make high speed photographs of the golf ball from which he could
deduce the axis and rate of rotation, and thereby all components of its angular
momentum, since these are compatible parts of a complete description of a
macroscopic spinning body.  But no corresponding possibility is available to
the quantum physicist: the different components of angular momentum of a
spin-half particle are \emph{incompatible}, and since trying to combine one
component with another yields a meaningless result, no measurement could
possibly determine the two values simultaneously.  And saying, ``I measured
$S_x=-1/2$ in this case; what would have been the result had I decided instead
to measure $S_z$?''  is to pose a tricky counterfactual question which easily
leads to misunderstanding.\cite{ntk12}

\section{Quantum Time Dependence}
\label{sct4}

If Schr\"odinger's (time-dependent) equation is deterministic, how is it
possible to introduce in a fundamental way a stochastic or probabilistic time
development in quantum theory?  Born's simple but ingenious idea\cite{br26} was
to use Schr\"odinger's equation to calculate probabilities.  The following
analogy may be helpful.  Classical Brownian motion of a particle modeled by a
Wiener process is random: the future behavior of the particle is not determined
by its present position or its past behavior. Nonetheless the probability
distribution density $\rho(\bld{r},t)$ for its position as a function of time
$t$ satisfied the \emph{deterministic} diffusion equation
\begin{equation}
  \pr\rho/\pr t = D\nabla^2\rho.
\label{eqn1}
\end{equation}
Why cannot one think of Schr\"odinger's equation in a similar way, as a
deterministic equation that generates probabilities?

One can, and in fact current textbooks do use Schr\"odinger's equation for this
purpose, but in a half-hearted and somewhat inconsistent way.  Following a
tradition that goes back at least to von Neumann,\cite{ntk07} the time
evolution of a quantum system is thought of as involving two distinct steps.
First one solves Schr\"odinger's equation to obtain a deterministic unitary
time development of the wave function, which tends to be thought of intuitively
as representing the ``real'' physical state of the microscopic quantum system.
Then the system of interest interacts with an external measuring apparatus,
resulting in a random process that leads to a situation in which the
measurement outcome, the only thing to which a probability can properly be
applied, is somehow associated with the state of the particle \emph{after} the
measurement has been completed.  The unsatisfactory nature of this approach
using wave function collapse has already been discussed in Sec.~\ref{sct2}.

Probabilities can be introduced in a more consistent and natural way by
following the route used in ordinary probability theory.\cite{ctk21} There the
first step is to introduce a \emph{sample space} of mutually-exclusive
\emph{events}, one and only one of which occurs in any particular experiment.
For example, if one rolls a die, the number of spots on the top face when it
comes to rest will be a number between 1 and 6; if 5 occurs, 3 does not occur,
etc.  The quantum counterpart is a set of mutually-orthogonal subspaces of the
Hilbert space whose projectors (orthogonal projection operators) form a
\emph{decomposition of the identity} $I$: a collection $\{P_j\}$ satisfying
\begin{equation}
  I = \sum_j P_j, \quad P_jP_k = \dl_{jk} P_j.
\label{eqn2}
\end{equation}
Note that $P_jP_k=P_kP_j$, so the properties are compatible; otherwise it would
not make sense to speak of one of them occurring rather than another; see the
discussion in Sec.~\ref{sct3}. The fact that $P_jP_k=0$ for $j\neq k$
corresponds to the properties being mutually exclusive: if one occurs the other
does not.  That the projectors sum to the identity means that one of them will
necessarily occur, or be true, at the time in question. An orthonormal basis
$\{\ket{\phi^j}\}$, $j=1,2\ldots$, in a finite-dimensional Hilbert space gives
rise to a a decomposition of the identity with $P_j = \dya{\phi^j}$.  Note that
real dice are quantum objects made up of atoms, hence describable (in
principle) using a large Hilbert space, and any visibly distinct states, such
as those with different numbers of spots on the top face, will correspond to
mutually-orthogonal projectors.  Thus \eqref{eqn2} works for both microscopic
and macroscopic systems, as one would expect, since the basic principles of
quantum mechanics apply to systems of any size.

A classical probabilistic description of a random (stochastic) process also
uses a sample space. In the case of a coin flipped three times it consists of
the 8 mutually exclusive possibilities, here called \emph{histories}, $HHH$,
$HHT$, $HTH$, \dots, $TTT$, where $H$ stands for ``heads'' and $T$ for
``tails.''  (Note that two histories are distinct elements of the sample space
if they differ at any of the three times.)  In the same way, in quantum
mechanics histories are sequences of quantum events at a succession of times,
each represented by a subspace (or its projector) of the quantum Hilbert space.
(For technical reasons it is convenient to represent histories as projectors on
tensor products of copies of the system's Hilbert space.\cite{ntk13}) The
behavior of a real coin made up of atoms can be described in quantum terms
using a suitable (large) Hilbert space, so the 8 mutually exclusive
possibilities of flipping it three times in a row also form a quantum sample
space or \emph{family} of histories.

Sample spaces are needed to make probabilistic reasoning precise, and while the
sloppy physicist's approach that ignores this is adequate for many purposes, in
quantum mechanics it leads to confusion. The first step in clearing up the
conceptual difficulties which have bothered Feynman and everyone else is to
introduce well-defined sample spaces for probabilities.  The second step is to
insist that \emph{incompatible} sample spaces not be combined, for the
combination will not make sense.  In the histories approach this is done by a
strict application of what is called the \emph{single framework
rule},\cite{ntk14} which asserts in essence that quantum probabilistic
reasoning must be carried out using a \emph{single} sample space.  Given two
\emph{compatible} quantum sample spaces this single space is easily constructed
from them by a process of refinement\cite{ntk15}, whereas if they
are incompatible the refinement does not exist.  Combining incompatible quantum
sample spaces in a way contrary to the single framework rule is at the heart of
most quantum paradoxes, and identifying the point at which this happens is the
key step in resolving (or, as I prefer to say, taming) such a paradox.

Once a sample space or family of histories has been defined, the next task is
assigning probabilities. For present purposes it suffices to consider a finite
sample space, so the probabilities are a collection of nonnegative numbers, one
for each history in the space, that sum to 1.  Probability theory as such
contains no rules for assigning these probabilities.  In
quantum theory Schr\"odinger's equation can be used to assign probabilities to
certain families of histories in a \emph{closed} or \emph{isolated} quantum
system (no interaction with something outside the system), the situation in
which Schr\"odinger's equation applies.  The simplest case involves only two
times $t_0$ and $t_1$, a single state $\ket{\psi_0}$ at time $t_0$, and an
orthonormal basis $\{\ket{\phi_1^j}\}$, $j=1,2,\ldots$, at $t_1$.  If the
Hamiltonian $H$ is independent of time the time evolution operator obtained by
integrating Schr\"odinger's equation is
\begin{equation}
  T(t',t) = e^{-i(t'-t)H/\hbar},
\label{eqn3}
\end{equation}
and the Born rule then gives
\begin{equation}
  \Pr(\phi_1^j\vb\psi_0) = |\mted{\phi_1^j}{T(t_1,t_0)}{\psi_0}|^2
\label{eqn4}
\end{equation}
as the conditional probability of $\ket{\phi_1^j}$ at time $t_1$ given
$\ket{\psi_0}$ at $t_0$.  The fairly obvious generalization (see \eqref{eqn2})
\begin{equation}
  \Pr(P_j\vb\psi_0) = \mte{\psi_0}{T(t_0,t_1)P_j T(t_1,t_0)}
\label{eqn5}
\end{equation}
of \eqref{eqn3} is also referred to as the Born rule. (Formulas
\eqref{eqn4} and \eqref{eqn5} apply if the Hamiltonian depends on time,
but then \eqref{eqn3} no longer gives the relationship between $T$ and $H$.)

Unlike those in quantum textbooks, the probabilities in \eqref{eqn4} and
\eqref{eqn5} do \emph{not} refer to outcomes of some \emph{external}
measurement, but to physical states \emph{inside} the closed system described
by the Hamiltonian used in \eqref{eqn3}. Born's rule is a fundamental law of
nature, on the same footing with Schr\"odinger's equation and equally
important. If one is interested in how a real measuring apparatus will interact
with a quantum system, one should include the apparatus itself as part of the
overall quantum system and then apply \eqref{eqn4} or \eqref{eqn5} to the
combination. Examples are discussed in Secs.~\ref{sct5} and \ref{sct6} below.
It is worth noting that $t_0$ may either precede $t_1$ or follow $t_1$. The
fundamental law for quantum probabilities, and its extensions (see below),
does not single out a sense of time. This important symmetry is entirely lost
sight of in the measurement-based approach to quantum theory, since
measurements are inherently irreversible (in the thermodynamic sense).

The right side of \eqref{eqn4} is often written as
$|\inpd{\phi_1^j}{\hat\psi_1}|^2$, where
\begin{equation}
  \ket{\hat\psi_1}=T(t_1,t_0)\ket{\psi_0}
\label{eqn6}
\end{equation}
is obtained from $\ket{\psi_0}$ by integrating Schr\"odinger's equation from
$t_0$ to $t_1$.  When used in this way $\ket{\hat\psi_1}$, which is typically
incompatible with the basis states $\{\ket{\phi_1^j}\}$, does \emph{not}
represent the physical reality of the quantum system at time $t_1$. It is
instead a mathematical construct, a pre-probability\cite{ntk16} used for
computing probabilities.  One could equally well compute them by starting with
each of the states $\ket{\phi_1^j}$ and integrating Schr\"odinger's equation in
the reverse direction from $t_1$ to $t_0$, making no reference whatsoever to
$\ket{\hat\psi_1}$.  For further discussion, see Sec.~9.4 of
Ref.~\onlinecite{gr02}.

Indeed, $\ket{\hat\psi_1}$ could be the infamous Schr\"odinger cat
state\cite{sch35}.  To discuss whether the cat is dead or alive, one should use
a framework, that is to say an orthonormal basis (or, to be more practical, a
decomposition of the identity) for which such concepts make sense, and then
compute probabilities. Since $\ket{\hat\psi_1}$ is a computational tool, it
requires no physical interpretation, and within the context of this framework,
it cannot be given a physical interpretation, for it is incompatible with the
sample space used to describe whether the cat is still alive. To be sure, one
could instead adopt a different, incompatible framework or orthonormal basis
that includes $\ket{\hat\psi_1}$ as one of its elements, in which case Born's
formula will tell us that it occurs with (conditional) probability 1. In this
second framework it makes no sense to ask whether the cat is dead or alive,
since the corresponding quantum properties are incompatible with
$\ket{\hat\psi_1}$.  In quantum mechanics, as in the case of Mount Shasta,
certain perspectives are useful for answering certain questions, and are not
useful for answering other questions.  The trouble with most treatments of
Schr\"odinger's cat is that they attempt to discuss its morbidity while
assuming that $\ket{\hat\psi_1}$ is its physical state, which makes no more
sense than talking about $S_z$ for spin-half particle whose $x$ component of
angular momentum is $+1/2$.

For a complete stochastic description of time development of a closed quantum
system it is necessary to go beyond the Born rule and provide formulas for
calculating probabilities of histories involving three or more times.  This
extension is not trivial, as consistent probabilities can only be assigned
if certain \emph{consistency} conditions are satisfied.  Discussing them here
would lead to a somewhat lengthy detour from our main theme, and as they are
treated in detail elsewhere\cite{ntk17}, we shall move
on to describe how consistent probability assignments within the context of
simple models can help dissipate quantum mysteries. 

\section{Toy Models}
\label{sct5}

A major difficulty in teaching quantum mechanics is that solving the
time-dependent Schr\"odinger equation is at best a time-consuming process, and
often cannot be done in closed form. This makes it difficult for students to
gain an intuitive understanding of what it involves.  The advent of computer
simulations with graphical output\cite{ntk51} is thus a welcome addition to the
repertoire of teaching tools.  However, these need to be supplemented by an
alternative approach using toy models, which, while somewhat unrealistic, have
the virtue that they can be worked out using a pencil on the back of the
traditional envelope.\cite{ntk21} 

\begin{figure}[h]
$$
\begin{pspicture}(-2.1,-0.6)(4.8,2.5) % (lower left)(upper right)
\newpsobject{showgrid}{psgrid}{subgriddiv=1,griddots=10,gridlabels=6pt}
		%PARAMETERS
\def\lwd{0.035} %Linewidth default
\def\lwb{0.10}  %Broad line
\def\rdot{0.1} \def\rodot{0.1} %Radius of closed, open dot
		%DEFAULTS
%\Huge
\psset{%unit=1.0cm, %1.0 cm is default
labelsep=2.0,
arrowsize=0.150 1,linewidth=\lwd}
		%OBJECTS
	% Dots
\def\cldot{\pscircle*(0,0){\rdot}}
\def\opdot{\pscircle[fillcolor=white,fillstyle=solid](0,0){\rodot}}
\def\dput(#1)#2#3{\rput(#1){#2}\rput(#1){#3}}
	% Labeled nodes
\def\clnr(#1)#2{\dput(#1){\cldot}{\rput[r](-0.2,0){$#2$}}} %clnr(0,1){a}
\def\clnl(#1)#2{\dput(#1){\cldot}{\rput[l](0.2,0){$#2$}}} %clnl(0,1){a}
\def\clnb(#1)#2{\dput(#1){\cldot}{\rput[b](0,0.2){$#2$}}} %clnb(0,1){a}
\def\clnt(#1)#2{\dput(#1){\cldot}{\rput[t](0,-0.2){$#2$}}} %clnt(0,1){a}
\def\opnr(#1)#2{\dput(#1){\opdot}{\rput[r](-0.2,0){$#2$}}} %opnr(0,1){a}
\def\opnl(#1)#2{\dput(#1){\opdot}{\rput[l](0.3,0){$#2$}}} %opnl(0,1){a}
\def\opnb(#1)#2{\dput(#1){\opdot}{\rput[b](0,0.2){$#2$}}} %opnb(0,1){a}
\def\opnt(#1)#2{\dput(#1){\opdot}{\rput[t](0,-0.2){$#2$}}} %opnt(0,1){a}
	% Arrows: right, left, up, down, upleft, upright
\def\rarr(#1){\rput(#1){\psline{->}(-0.2,0)(0,0)}}
\def\larr(#1){\rput(#1){\psline{->}(0.2,0)(0,0)}}
\def\uarr(#1){\rput(#1){\psline{->}(0,-0.2)(0,0)}}
\def\ularr(#1){\rput(#1){\psline{->}(0.2,-0.2)(0,0)}}
\def\urarr(#1){\rput(#1){\psline{->}(-0.2,-0.2)(0,0)}}
\def\darr(#1){\rput(#1){\psline{->}(0,0.2)(0,0)}}
		%DIAGRAM
\psline{->}(0,2)(4.5,2)
\psline{-<}(0,0)(4.5,0)
\psline(0,0)(-1,1)(0,2)(0,0)
\opnb(0,2){1}\opnb(1,2){2}\opnb(2,2){3}\opnb(3,2){4}\opnb(4,2){5}
\rarr(0.6,2)\rarr(1.6,2)\rarr(2.6,2)\rarr(3.6,2)
\opnt(0,0){-1}\opnt(1,0){-2}\opnt(2,0){-3}\opnt(3,0){-4}\opnt(4,0){-5}
\larr(0.4,0)\larr(1.4,0)\larr(2.4,0)\larr(3.4,0)
\ularr(-.6,0.6)\urarr(-0.4,1.6)\uarr(0,1.1)
%\psbezier{->}(-1,1)(-1,1.7)(-2,1.7)(-2,1)
%\psbezier(-1,1)(-1,0.3)(-2,0.3)(-2,1)
%\pscircle(-1.5,1){0.5}
%\psarc(-1.5,1){0.5}{-180}{180}
\psarc{->}(-1.5,1){0.5}{-180}{190}
\opnl(-1,1){0}
\rput[rb](-.2,2.2){$m=$}
\rput[rt](-.3,-0.3){$m=$}
\psline{->}(2.2,1.6)(4.7,1.6)
\clnt(1.2,1.6){0}\clnt(2.2,1.6){1}\clnt(3.2,1.6){2}\clnt(4.2,1.6){3}
\rarr(2.8,1.6)\rarr(3.8,1.6)
\rput[br](1.0,1.1){$n=$}
\rput[l](-1.9,1){$\al$}\rput[rb](-.5,1.5){$\bt$}
\rput[l](0.1,0.6){$\al^*$}\rput[rt](-0.5,0.4){$-\bt^*$}
%\showgrid %Comment this line to remove grid
\end{pspicture}
$$
\caption{Toy model of particle decay with detector.  The sites $m$ refer to the
particle and $n$ to the pointer of the detector.}
\label{fgr2}
\end{figure}
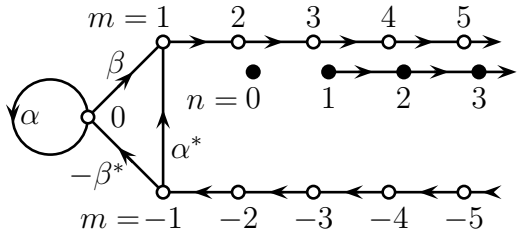

The basic idea is to discretize time so that it advances in integer steps, and
the time development operator in \eqref{eqn3} takes the form of an integer
power
\begin{equation}
  T(t',t) = T^{t'-t}
\label{eqn7}
\end{equation}
of some very simple unitary operator $T$, typically one representing a hopping
motion of one or more particles.  For example, $T=S$,
where
\begin{equation}
  S\ket{m} = \ket{m+1},\quad S\ket{M} = \ket{-M}
\label{eqn8}
\end{equation}
is a shift operator moving a particle from a lattice site or node or ``box'' at
site $m$, where $m$ is an integer, to the next site. The periodic boundary
condition in \eqref{eqn8} ensures that $S$ is a unitary operator on the
finite-dimensional Hilbert space with orthonormal basis $\{\ket{m}\}$, $-M\leq
m \leq M$, where $M$ can be as large as one wants; typically much larger than
the times of interest.  Figure~\ref{fgr2} shows a modification in which
\eqref{eqn8} holds except for $m=0$ and $-1$, for which
\begin{equation}
  S\ket{0} = \al\ket{0} + \bt\ket{1},\quad  
  S\ket{-1} = -\bt^*\ket{0} + \al^*\ket{1},
\label{eqn9}
\end{equation}
with $|\al|^2+|\bt|^2=1$.
One can think of this as a simple model of a decaying system: an alpha
particle initially inside a nucleus at $\ket{m=0}$ eventually escapes to $m=1$
and then keeps moving.  The unitary time development of an initial state
$\ket{\psi_0}=\ket{0}$ at $t=0$ leads to
\begin{equation}
  \ket{\psi_t} = T^t \ket{\psi_0} = \al^t\ket{0} + \bt\left[ \al^{t-1}\ket{1} +
    \al^{t-2}\ket{2}+\cdots +\ket{t}\right]
\label{eqn10}
\end{equation}
for $0<t<M$. Born's rule gives $|\al|^{2t}$ for the probability that the
initial state has not yet decayed. This decreases exponentially with $t$, as
one might expect. 

A way to make this model a useful tool for
dissipating quantum mysteries is to add a toy detector, thought of as the
pointer on a toy measuring apparatus, with states labeled $n$ in
Fig.~\ref{fgr2}.  Let
\begin{equation}
  S'\ket{n}=\ket{n+1}, \text{ except } S'\ket{0}=\ket{0} \text{ and }
  S'\ket{-1}=\ket{1},
\label{eqn11}
\end{equation}
be the corresponding shift operator, and again assume a periodic boundary
condition $S'\ket{N} = \ket{-N}$. The total time development operator on the
tensor product of the particle and pointer Hilbert spaces is
\begin{equation}
  T = (S\ot I) R (I\ot S'),
\label{eqn12}
\end{equation}
where $R(\ket{m}\ot\ket{n}) = \ket{m}\ot\ket{n}$ is the identity $I\ot I$
except for
\begin{equation}
  R(\ket{2}\ot\ket{0}) = \ket{2}\ot\ket{1}, \quad
  R(\ket{2}\ot\ket{1}) = \ket{2}\ot\ket{0}.
\label{eqn13}
\end{equation}
If the detector pointer is initially at $n=0$ in its ``ready'' state and the
particle arrives at $m=2$, the effect of $T$ is to kick the pointer to $n=1$,
after which it continues moving. At the same time the particle continues on to
$m=3$, as it would have done in the absence of the detector.

Unitary time development of an initial state
$\ket{\Psi_0}=\ket{m=0}\ot\ket{n=0}$ to a time $t\geq 3$ results in
\begin{multline}
  \ket{\Psi_t}= T^t \ket{\Psi_0} = \left[ \al^t\ket{0} + \bt\al^{t-1}\ket{1}
 +\bt\al^{t-2}\ket{2}\right]\ot \ket{0}\\
   + \bt\left[ \al^{t-3}\ket{3}\ot\ket{1}
 +\al^{t-4}\ket{4}\ot\ket{2} +\cdots+ \ket{t}\ot\ket{t-2}\right].
\label{eqn14}
\end{multline}
Notice that in this expression the detector is in a superposition of different
pointer positions, so we have the toy analog of a Schr\"odinger cat---a
Schr\"odinger kitten?  A useful physical interpretation is obtained by using
the Born rule \eqref{eqn4} at time $t_1=t$, with the orthonormal basis
$\{\ket{m}\ot\ket{n}\}$, i.e., both particle and pointer are at definite
locations.  If we think of $\ket{\Psi_t}$ as a pre-probability, the analog of
$\ket{\hat\psi_1}$ in \eqref{eqn6}, then the probability that the particle is
at $m$ and the pointer at $n$ at time $t$ is just the absolute square of the
corresponding coefficient on the right side of \eqref{eqn14}.  This joint
probability distribution $\Pr(m,n)$ has exactly the same properties and the
same physical interpretation as in ordinary probably theory.  In particular, we
can use it to compute the conditional probabilities $\Pr(m\vb n)$, and from
them deduce that if the pointer is at $n=0$, then $m\leq 2$, i.e., the alpha
particle is still in the nucleus or on its way to the detector; whereas if the
pointer is at some $n>0$, the particle is at the location $m=n+2$, as one would
expect if the particle triggered the detector while hopping from $m=2$ to
3. Quantum mechanics does not say which of these mutually exclusive and
physically reasonable possibilities is actually the case, but only provides
probabilities.

This simple example, in which the measuring device is part of the total quantum
system, is useful in countering a number of misleading ideas that students
unfortunately pick up while taking elementary (and more advanced) quantum
courses: that particles (and pointers) can be in two places at the same time,
that quantum mechanics necessarily leaves everything in a fog, that there are
magical long-range influences, etc. Note in particular how wave function
collapse is not needed when probabilities are introduced in a consistent way
into quantum theory.  Removing wave function collapse from textbooks and
replacing it with conditional probabilities would be a significant step towards
improving students' understanding of quantum mechanics.

The preceding discussion might tempt one to conclude that if at $t=5$ the
pointer is at $n=1$, then at $t=2$ the particle was at $m=1$.  This conclusion
is correct, but cannot be justified on the basis of the Born rule alone, as it
involves probabilistic reasoning applied to a closed quantum system at 3
different times: the initial state at $t=0$, the pointer position at $t=5$, and
the particle position at $t=3$.  One must use an appropriate extension of the
Born rule and check for consistency.\cite{ntk22}  We will give
another example in the next section of how measurement outcomes can be used to
infer properties of a measured system before the measurement took place.

\section{Measurements Reconsidered}
\label{sct6}

Measurement apparatus is essential for experiments exploring the quantum
properties of microscopic systems, for it amplifies very small effects and
makes them visible or audible or otherwise evident in macroscopic effects
accessible to human beings.  Thus it is very important to understand how the
apparatus works, and how its macroscopic output is related to the microscopic
input.  Does the process introduce noise, and if so how much?  Is the output
influenced by extraneous effects?  These questions can be studied to some
extent by carrying out experimental tests.  But an important theoretical
component goes into such analyses, and in this respect measurement-based
quantum mechanics as found in the textbooks is inadequate.  It is hard to
analyze real measurements when the very concept of a measurement is considered
as axiomatic, and thus unanalyzable in quantum terms.  Introducing
probabilities in a consistent way makes it possible, in principle, to analyze
real apparatus in a completely quantum mechanical way.  Future
experimentalists, and theorists who give them advice, need to know that a
consistent approach to these questions exists, that it does not depend upon
dubious ideas like wave function collapse, and that it supports many of the
general intuitions which experimental physicists have about measuring
apparatus, such as the fact that if there is a collimator between source and
particle detector, then on its way to the detector the particle has to pass
through the hole in the collimator.  At the same time it places limits on that
intuition, and indicates places at which it will break down and caution needs
to be observed.  This article is not the place to go into details, but the most
essential ideas can be explained in terms of a simple example, a somewhat
idealized and modernized version of the famous Stern-Gerlach
measurement.\cite{gs22} This will show how the measurement-based approach of
textbooks can be unhelpful and misleading even when it is in some respects
correct, and how to replace it with something more useful.

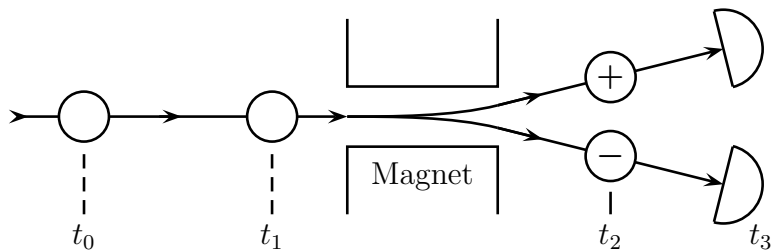
\begin{figure}[h]
$$
\begin{pspicture}(-4.6,-1.8)(5.7,1.5) % (lower left)(upper right)
\newpsobject{showgrid}{psgrid}{subgriddiv=1,griddots=10,gridlabels=6pt}
		%PARAMETERS
\def\lwd{0.035} %Linewidth default
\def\tkw{0.1} % Half width of tick
		%DEFAULTS
%\Huge
\psset{%unit=1.0cm, %1.0 cm is default
labelsep=2.0,
arrowsize=0.150 1,linewidth=\lwd}
		%OBJECTS
	% Double put. \dput(x,y){First}{Second}
\def\dput(#1)#2#3{\rput(#1){#2}\rput(#1){#3}} 
	% Ticks \htick = horizontal tick for vert axis; \vtick = vertical tick
\def\vtick{\psline(0,-\tkw)(0,\tkw)}
	% Blanking circle \rput(x,y){\circ}
\def\circ{% \rput(x,y){\circ}Enter radius value in {}
\pscircle[fillcolor=white,fillstyle=solid]{0.35}}% END circ
		%DIAGRAM
	%Particle lines
\psline{>->}(-4.5,0)(0,0)
\psbezier(0,0)(0.6,0)(1.4,0)(2,0.15)
\psline{->}(2,0.15)(5,0.9)
\psbezier(0,0)(0.6,0)(1.4,0)(2,-0.15)
\psline{->}(2,-0.15)(5,-0.9)
% Additional arrows
\psline{->}(-2.5,0)(-2.2,0)
\psline{->}(2.0,0.15)(2.6,0.30)
\psline{->}(2.0,-0.15)(2.6,-0.30)
	%Wave packet circles
\rput(-3.5,0){\circ}\rput(-1.0,0){\circ}
\dput(3.5,0.525){\circ}{$\bld{+}$}
\dput(3.5,-0.525){\circ}{$\bld{-}$}
	%Stern-Gerlach Apparatus
\psline(0,1.3)(0,0.4)(2,0.4)(2,1.3)
\psline(0,-1.3)(0,-0.4)(2,-0.4)(2,-1.3)
\rput(1,-.8){Magnet}
	%Detectors
\psline(4.875,1.4)(5.125,0.4)
\psarc(5.0,0.9){0.522}{-75.96}{104.04}
\psline(4.875,-1.4)(5.125,-0.4)
\psarc(5.0,-0.9){0.522}{-104.04}{75.96}
	%Times
\rput[B](-3.5,-1.7){$t_0$}
\rput[B](-1.0,-1.7){$t_1$}
\rput[B](3.5,-1.7){$t_2$}
\rput[B](5.5,-1.7){$t_3$}
\psline[linestyle=dashed](-3.5,-0.5)(-3.5,-1.3)
\psline[linestyle=dashed](-1.0,-0.5)(-1.0,-1.3)
\psline[linestyle=dashed](3.5,-1)(3.5,-1.3)
%\showgrid %Comment this line to remove grid
\end{pspicture}
$$
\caption{Stern-Gerlach apparatus separating particles into $S_z=\pm1/2$ beams,
which are then detected. }
\label{fgr3}
\end{figure}

Figure~\ref{fgr3} shows the well-known schematic diagram: a stream of spin-half
particles enter on the left and are separated into two outgoing beams: the
upper one corresponding to $S_z=+1/2$ and the lower to $S_z=-1/2$, for a
magnetic field gradient in the $z$ direction.  That is, if at time $t_1$ just
before entering the apparatus the spin state is $S_z=+1/2$, the particle will
emerge in the upper beam, and can be detected by the upper detector.
Similarly, if $S_z=-1/2$ the particle will emerge in the lower beam and be
detected there.  We suppose that the magnetic field is negligible at and to
the left of $t_1$ in Fig.~\ref{fgr3}.

What will happen if the particle is prepared, via some previous apparatus, so
that it is in in a state with $S_x=+1/2$ at a time $t_0< t_1$? Since $S_x=+1/2$
is a linear superposition of the $S_z=+1/2$ and $S_z=-1/2$ states with equal
amplitude, the standard (correct) answer is that it will be \emph{detected}
with probability $1/2$ in the upper and probability $1/2$ in the lower beam.
Suppose it has been detected by the upper detector, as indicated by a pointer
on that device, at time $t_3$.  Was the particle in the upper beam at time
$t_2$, after leaving the field gradient but before detection?  Experimental
physicists will tend to answer that it was, for otherwise they will have
difficulty designing equipment, thinking about errors, etc.  Theoretical
physicists trained in the usual textbook approach may disagree, for they think
of the original spin superposition as developing unitarily into a superposition
of two wave packets, one traveling upwards and one downwards after the atom
leaves the field gradient.  (Let us assume the vacuum is good enough that
decoherence from collisions does not complicate matters.)  And what can one say
about the spin state of a particle at the earlier time $t_1$ if it is later
detected by the upper detector?

All of these questions have reasonable answers if one abandons the measurement
approach and instead introduces microscopic quantum probabilities on
appropriate sample spaces, that is, consistent families of quantum
histories. The key issue is the choice of sample space, for in a situation of
this sort there are several \emph{incompatible} alternatives. We will consider
various possibilities, always assuming as given data an initial $S_x=+1/2$ spin
state and detectors in the ready state at time $t_0$, and that at time $t_3$ it
is the upper detector that has been triggered by the arrival of the particle.
Note that the detectors are here thought of as part of a large closed quantum
system that also includes the particle.

A first consistent family $\FS_a$ can be represented, using the notation
employed in Ref.~\onlinecite{gr02}, in the form
\begin{equation}
  \begin{array}{l c\csp c\csp c\csp c\csp c\dsp c\csp c\csp c} %9 cols, 8 &
  &&&&&& u & \od & U\\
 \rbx{\FS_a\text{ : }} & \rbx{x^+} & \rbx{\od} & \rbx{I} & 
 \rbx{\od} &\rbx{\lbrc} & l & \od  & L \\
 & t_0 && t_1 &&& t_2 && t_3
   \end{array}
\label{eqn15}
\end{equation}
Here each letter represents a projector in a history associated with the four
successive times $t_0<t_1<t_2<t_3$ indicated on the lower line, and the $\od$
symbols can for present purposes be thought of as commas separating the
projectors at successive times.\cite{ntk22b} In particular, $x^+$ at the time
$t_0$\cite{ntk22a} means $S_x=+1/2$, the identity $I$ at $t_1$ indicates that
no information is being provided about the state of the particle at this time
(in contrast to \eqref{eqn18} and \eqref{eqn20} below), $u$ and $l$ at $t_2$
signify that the particle is in the upper and lower path, respectively, while
$U$ and $L$ are projectors corresponding to the upper and lower detector,
respectively, having detected the particle at $t_3$.  One can think of
\eqref{eqn15} as a shorthand for two histories, $x^+\ot I \ot u\ot U$ and
$x^+\ot I \ot l\ot L$, with the curly brace indicating that they are identical
up to the time $t_1$.
The extended Born rule\cite{ntk22c} assigns a probability of 1/2 to each of the
two histories in \eqref{eqn15}.  The conditional probabilities
\begin{equation}
  \Pr(u_2\vb U_3) =1,\quad \Pr(l_2\vb U_3) =0,
\label{eqn16}
\end{equation}
where the subscripts refer to times $t_2$ and $t_3$, follow at once from the
fact that there is only one history in $\FS_a$ for which the upper detector
triggers, and in that history $U$ is preceded by $u$, not $l$.  What
\eqref{eqn16} tells us is that if the upper detector triggers, one can be
certain that at the earlier time $t_2$ the particle was following the upper and
not the lower path. So the experimentalist is right.

But there is also a second consistent family
\begin{equation}
  \begin{array}{l c\csp c\csp c\csp c\csp c\csp c\csp c \dsp c} %9 cols, 8 &
  &&&&&& & & U\\
 \rbx{\FS_b\text{ : }} & \rbx{x^+} & \rbx{\od} & \rbx{I} & 
 \rbx{\od} &\rbx{c}& \rbx{\od} &\rbx{\lbrc} & L 
   \end{array}
\label{eqn17}
\end{equation}
where the times are the same as in \eqref{eqn15}.  Here $c$ at $t_2$ is a
projector onto the coherent superposition of states that evolve from the
initial state with $S_x=+1/2$, and since it is found in both histories, it
occurs with probability 1, just as the theoretician supposed.
Since both $\FS_a$ and $\FS_b$ are consistent families, the conclusions of a
probabilistic analysis applied using just one of them while disregarding the
other will be correct.  However, the families are incompatible, and so these
conclusions cannot be combined.  One cannot say that at time $t_2$ the particle
is both in a superposition state $c$ AND that it is moving on the upper
trajectory $u$, for that would be meaningless in the same way that ``$S_x=+1/2$
AND $S_z=+1/2$'' makes no sense.  Note that incompatibility, the fact that the
two families cannot be combined, does \emph{not} mean that one is ``wrong'' and
the other is ``right.''  Seeking some law of nature which ``chooses'' one
rather than the other is to misunderstand the nature of quantum
descriptions. It is the physicist who chooses which description to use,
depending upon the sort of question he is asking, while noting that only
descriptions compatible with the desired information will be useful for this
purpose.  Remember Mount Shasta.

Thus far we have said nothing about the spin state of the particle at the time
$t_1$ when it is just about to enter the field gradient; both $\FS_a$ and
$\FS_b$ contain a noncommittal $I$ at $t_1$.  It is again useful to consider
two different consistent families.  In
\begin{equation}
  \begin{array}{l c\csp c\csp c\dsp c\csp c\csp c\csp c\csp c} %9 cols, 8 &
  &&&&  
   z^+  & \od  & u & \od & U\\
 \rbx{\FS_c\text{ : }} & \rbx{x^+} & \rbx{\od} & \rbx{\lbrc}  & 
   z^-  & \od  & l & \od & L
   \end{array}
\label{eqn18}
\end{equation}
one can talk about $S_z$ at $t_1$: the projectors $z^+$ and $z^-$ correspond to
$S_z=\pm 1/2$. Once again there is only one history that terminates in
$U$, and therefore
\begin{equation}
  \Pr(z^+_1\vb U_3) =1,\quad \Pr(z^-_1\vb U_3) =0.
\label{eqn19}
\end{equation}
That is, one can be sure that if the upper detector detected the particle,
$S_z$ had the value $+1/2$, not $-1/2$, at the earlier time $t_1$. This is what
one would expect if the total apparatus, which consists of field gradient
followed by detectors, functions as designed, as a device to measure the $z$
component of the spin of a spin-half particle.

In the second consistent family
\begin{equation}
  \begin{array}{l c\csp c\csp c\csp c\csp c\csp c\csp c \dsp c} %9 cols, 8 &
  &&&&&& & & U\\
 \rbx{\FS_d\text{ : }} & \rbx{x^+} & \rbx{\od} & \rbx{x^+} & 
 \rbx{\od} &\rbx{I}& \rbx{\od} &\rbx{\lbrc} & L 
   \end{array}
\label{eqn20}
\end{equation}
it is $S_x$ that makes sense at $t_1$, and $S_x=+1/2$ occurs with probability
1.  This family is of no use in deciding whether the measuring apparatus is
functioning properly, since that question makes reference to $S_z$ at $t_1$ and
not $S_x$, but it could provide a check on whether the region traversed by the
particle during the interval from $t_0$ to $t_1$ was free of magnetic fields,
as we have supposed.  Of course, $\FS_d$ is incompatible with $\FS_c$, so it
makes no sense to combine the probability 1 inferences obtained by using them
separately.  (Incidentally, in $\FS_d$ one could replace the $I$ at time $t_2$
with the pair $u$ and $l$, as in \eqref{eqn18}. The result would be a family
$\FS'_d$ which would serve equally well for the matters we have been
discussing. Likewise, in $\FS_c$ one could replace $u$ and $l$ at $t_2$ with
$I$.)

The following conceptual difficulty can arise when using the family
$\FS_c$. How can it be that $S_x=+1/2$ at $t_0$ (as an initial datum) and
$S_z=+1/2$ at $t_1$ (with probability 1) if there is no magnetic field acting
on the particle during the time interval between $t_0$ and $t_1$, and thus no
torque which could have caused the spin direction to precess from $+x$ to $+z$?
This problem arises from a misleading mental picture of a spin-half particle in
the state $S_x=+1/2$.  One tends to think of it as a little gyroscope with its
axis of rotation lined up precisely along the $+x$ axis, and if at a later time
the gyroscope axis is in the $+z$ direction, this must have come about through
the application of a torque.  But a gyroscope has $y$ and $z$ components of
angular momentum equal to 0 if its axis is in the $x$ direction, whereas for a
spin-half particle these other components are undefined when $S_x=+1/2$. A
better, less misleading image is to think of $S_x=+1/2$ as resembling a
gyroscope with its axis in a random direction, i.e., random $y$ and $z$
components of angular momentum, subject only to the constraint that the $x$
component is fixed.  Then even if the gyroscope is not subject to a torque,
there is no reason why its $x$ component of angular momentum cannot be positive
at one time and its $z$ component positive at a later time.  Classical images
of some sort are probably essential in quantum physics, since they help us
organize intuitive knowledge, and they always mislead to some extent.  But some
mislead less than others, as shown by this example.

One can continue the discussion of the Stern-Gerlach experiment using
additional families of histories which combine information about a spin
component at $t_1$ with information about a position or superposition of
positions at $t_2$, but the preceding suffices for making the main points.
Families $\FS_b$ and $\FS_d$ correspond in a rough sense to the viewpoint of
von Neumann and the typical textbook, in which unitary time development
persists up until the last instant before the final measurement, meaning the
amplification of a microscopic signal to a macroscopic level, takes place.
Thus they show that the textbook approach makes a certain amount of
sense. However, the conclusions we reached using $\FS_b$ and $\FS_d$ are based
on the systematic use of fundamental principles of quantum dynamics applied to
a closed quantum system, not on anything specific to a measurement, and
standard probabilistic reasoning, not guesswork or arm waving.

On the other hand, $\FS_a$ and $\FS_c$ provide the sort of information needed
by someone designing a quantum measuring apparatus, or analyzing how it
functions.  The key point is that such an analysis in quantum terms is only
possible if the relevant properties of the measured system at a time
\emph{before} the measurement takes place are part of the quantum description.
This is not so in the usual textbook approach, which is defective not in that
it is wrong---as we have seen, it can be justified to some extent by using
families like $\FS_b$ and $\FS_d$---but in that equally valid alternatives for
discussing quantum time development are never mentioned, and the student is
left with the incorrect idea that quantum measurements really do not measure
anything, they just cause the great smoky dragon\cite{wh83b} to collapse.

\section{Practical Considerations}
\label{sct7}

For a period of ten years I have been teaching various advanced undergraduate
and beginning graduate quantum mechanics courses, and courses in quantum
information, using the new perspective in which quantum mechanics is based on
probabilistic laws of universal validity, with measurements being only one of
the applications.  The reaction of students has generally been positive, though
there are always signs of shock when I tell them that by the end of the course,
and provided they do their homework, they will understand (some aspects of)
quantum mechanics better than Feynman did.  Homework and examinations results
indicate that they understand this material about as well, or as badly, as
other topics in such courses, but there have been no follow-up studies to see
what they have retained a year later.

How long does it take to present the new ideas?  Longer than the material they
replace, but not enormously so.  Courses at the advanced undergraduate and
beginning graduate level typically devote a certain amount of time to
introducing fundamental quantum concepts; defining a quantum Hilbert space,
Dirac notation, tensor products; introducing Schr\"odinger's equation and a
probabilistic interpretation of the formalism; and examples illustrating all of
these. Before moving on to angular momentum, the hydrogen atom, scattering, and
so forth.  It is in the first part that changes are most needed, and my
experience suggests that the revised version requires about six weeks total (of
a fourteen week semester) in an introductory graduate course; perhaps one or
two more than if one follows the older approach.  The new material includes a
proper discussion of quantum incompatibility; histories and consistency
conditions; the toy models needed to provide illustrations; and a one hour
introduction to probability theory for students who have not yet had a course
in that subject.  Along the way the students learn how to deal with the double
slit paradox, or the equivalent using a Mach-Zehnder interferometer, in a
reasonable way. Resolving the Einstein-Podolsky-Rosen problem\cite{epr35}
without invoking long range influences requires less than one additional class
period if the foundations have been properly laid.  There is no need to
consider Bell's inequality,\cite{bl64} though this can serve as a useful
illustration of what goes wrong when one tries to import classical ideas into
the quantum world.

What gives the students the most difficulty?  Quantum incompatibility.  The
problems they face are analogous to those encountered when first studying
relativity, only worse: habits of classical reasoning lie closer to the soul of
the apprentice physicist than does the notion of temporal simultaneity.
However, just as students are capable of learning that putting $x$ to the left
of $p$ in quantum theory does not yield anything like the classical $xp$, they
can also learn its logical counterpart, especially if one starts with the
simple case of spin half.  Next in order of (decreasing) difficulty come
consistency relations.\cite{ntk23} Followed by probability theory in the case
of students who have never been exposed to its formal structure, nor dealt
with simple stochastic processes.\cite{ntk24} Fortunately, in an introductory
quantum course one can get by with finite sample spaces and finite-dimensional
Hilbert spaces, with only some talk about their infinite counterparts, so the
formal mathematics is not very difficult.

A different kind of conceptual barrier can be present, especially for graduate
students who in previous courses taught by respected teachers have learned
the measurement-based approach to quantum mechanics with wave function
collapse, etc., while never becoming aware of its many shortcomings and
inconsistencies.  It is then hard to persuade them to pay serious attention to
something which appears contrary to what they think of as quantum
orthodoxy. Another objection that is raised, again primarily by graduate
students, is that they are being required to learn esoteric material about
quantum foundations, rather than how to do calculations that will aid them in
passing exams and preparing for research.  The fact that students are often
hesitant to express these reservations openly to the teacher makes it harder to
deal with them.  When countering prejudices of this type I think it not
inappropriate to point out that Feynman, who knew how to do calculations better
than most of us, was quite forthright in admitting that he did not understand
quantum mechanics as formulated in the traditional way, and that anecdotal
evidence suggests he was impressed by the new ideas when he first heard them
shortly before his untimely death.\cite{gh99}

What material is available for a course taught from the new point of view?  No
textbook, so far as I know, has incorporated the new ideas.\cite{ntk25} There
are two monographs by Omn\`es,\cite{ctk11} of which the second is simpler.
My own book\cite{gr02} is simpler still, and can be used as a supplement to a
regular textbook.  The most crucial chapters are available on the Internet,
along with a small number of exercises.

\section{Conclusion}
\label{sct8}

I have argued that the treatment of quantum probabilities found in textbooks,
where they are introduced in connection with measurement outcomes, is a major
source of conceptual difficulties for students trying to learn the subject.
And that modern developments in our understanding of quantum mechanics make it
possible to do a much better job, through a systematic and coherent
introduction of microscopic probabilities as a fundamental part of the
theory.  Measurements can then be understood as particular examples of quantum
processes, not as something fundamentally different, and can be shown to reveal
something about the measured system before the measurement took place.  Wave
function collapse can be assigned to the trash can of outmoded ideas, replaced
by a consistent use of conditional probabilities.  Furthermore, such an
approach is not beyond the grasp of students, especially when explained with
the aid of toy models that allow them to understand the fundamentals of quantum
dynamics without becoming entangled in the technical difficulties of solving
Schr\"odinger's equation.  As a consequence, students can now begin to
understand those aspects of quantum mechanics that Feynman found so difficult.

Even the reader who thinks these arguments have merit may well ask, and
properly so, whether it is really worthwhile replacing the traditional approach
embodied in standard textbooks with something newer.  Has not the older
approach, whatever its flaws, allowed several generations of physicists to
carry out excellent research?  Have not the textbooks been written by authors
with considerable pedagogical skill?  Indeed, are the conceptual gaps, which
even textbook writers themselves have sometimes acknowledged,\cite{ll01} all
that serious?  Do not good physicists, whether engaged in theory or experiment,
eventually develop the sort of intuition which allows them to work around
deficiencies in their courses?  Do not our present courses at least teach
students how to calculate things in agreement with experiment?

While sympathetic with such concerns, I must ask: Is it our primary goal to
impart calculational skills to our students?  No doubt this is one of the
things we aim to do.  The engineering student who can successfully apply the
formula numbered 37 in his freshman mechanics text to a physics problem will
succeed later, we hope, in applying the right formula from the appropriate
engineering handbook to some design problem.  The difficulty comes in
situations in which formula 37 is no longer applicable, or perhaps one is not
sure whether it applies, or maybe it is necessary to make some approximations,
and good judgment is needed as to whether these are appropriate, etc.  There
are lots of reasons why when we teach classical mechanics we want our students
not only to know the formulas in the blue boxes, but to imbed them in a real
understanding of the deeper principles of the subject.  If this is so, should
our goal in the case of quantum mechanics be different?

To be sure, in any discipline of physics one eventually arrives at principles
which in our present state of knowledge cannot be explained in terms of
anything more fundamental.  At that stage we have to stop and recognize that
there is a limitation to our understanding, there are things that simply have
to be accepted on faith, hopefully supported by the fact that they have been
shown to work in a large number of circumstances.  Especially when we have a
consistent and coherent framework for some subject there is no reason to
apologize, even when we know it is at best an approximation to the real world.
Classical electricity and magnetism has this character.  It is approximate (the
real world is quantized) and there are always some loose ends to be understood
better, but overall it is satisfactory, and we teach it with confidence to our
students.

The situation in quantum mechanics, as reflected in current textbooks, is
very different.  A significant contribution of decades of research in quantum
foundations has been to remind the community that quantum mechanics as
traditionally taught contains all sorts of unresolved problems and paradoxes
that cast serious doubt on its coherence as an intellectual discipline.  These
issues were often ignored in older textbooks, but newer ones feel obliged to
devote at least a few pages to Einstein-Podolsky-Rosen and similar things.
This acknowledges, in an indirect way, that the system being taught has serious
flaws, and in this respect textbook writers are at last catching up to what
Feynman was saying in 1964.  Is this flawed approach what we want to pass on to
our students, or should we aim for something better?

	\section*{Acknowledgments}
Discussions with and comments by E. Gerjuoy and C. Singh are gratefully
acknowledged.  The research described here received support from the National
Science Foundation through Grant PHY-0456951.


\begin{thebibliography}{11}
\bibitem{ntk01}
%
Different isomers decay at
different rates, but that can be understood on the basis of different internal
quantum states. Given a particular unstable isomer, we have no way to separate
individual nuclei on the basis of their future longevity.
\bibitem{ntk02}
%
Believers in these influences often concede that they cannot be used to
transmit information, which is the same as saying that no experimental test
will ever reveal their presence. In fact, a proper quantum analysis using
consistent microscopic probabilities shows that such influences are not needed
when one uses conditional probabilities to analyze the situations in which they
supposedly arise.  See the discussion in Chs.~23 and 24 of
Ref.~\onlinecite{gr02}.
\bibitem{ntk18}
See Chs.~20 through 25 of Ref.~\onlinecite{gr02} for a number of examples of
how paradoxes can be dealt with by a consistent use of probabilities.
\bibitem{ntk06}
%
For an
infinite-dimensional Hilbert space one requires that the subspace be closed.
\bibitem{ntk12} 
%
Counterfactual reasoning of the sort ``If it had been the case that\dots
then\dots'' is a particularly dangerous alligator.  See Ch.~19 of
Ref.~\onlinecite{gr02} for an explanation of how to do some forms of
counterfactual reasoning in a consistent way in the quantum context, and later
chapters for examples of what happens when one does it inconsistently.
\bibitem{ntk21}
Numerous examples are given in Ref.~\onlinecite{gr02}.
\bibitem{ntk22}
The process is not difficult, and is
carried out for various toy models in Chs. 12 and 13 of Ref.~\onlinecite{gr02}.
\bibitem{ntk22b}
The symbol $\ot$ is used in Ref.~\onlinecite{gr02} as a special form of
tensor product symbol $\ot$; see the discussion in Sec.~8.3.
\bibitem{ntk22a} 
%
The projector at $t_0$ should include, along with the spin state of the
particle, its initial spatial wave function and the initial (untriggered) state
of the two detectors, but as these are exactly the same in all the histories we
shall consider, and irrelevant to our discussion, there is no harm in omitting
them from the notation used in \eqref{eqn15} and later.
\bibitem{ntk23} 
%
I find it best when introducing the subject to use the approach in Sec.~11.6 of
Ref.~\onlinecite{gr02}, rather than the general formulation of Ch.~10; the
former suffices for most purposes.
\bibitem{ntk24}
%
See Ref.~\onlinecite{br02} for some of the difficulties students have with
probabilities.
\bibitem{ntk25} 
%
Anyone interested in such a project is welcome to contact me.
\end{thebibliography}
\end{document}